\pdfoutput=1

\documentclass[11pt]{article}

\newif\ifarxiv
\arxivtrue 

\ifarxiv
    \usepackage{acl}
\else
    \usepackage[review]{acl}
\fi
\usepackage{csquotes}
\usepackage{times}
\usepackage{latexsym}
\usepackage{amsmath}
\usepackage[T1]{fontenc}

\usepackage[utf8]{inputenc}

\usepackage{microtype}

\usepackage{inconsolata}
\usepackage{enumitem}
\usepackage{array}
\usepackage[capitalise]{cleveref}
\usepackage{listings}
\usepackage{graphicx}
\usepackage{longtable}
\usepackage{booktabs}
\usepackage{soul,xcolor}
\usepackage{longtable}
\sethlcolor{yellow}
\usepackage{subcaption}
\usepackage{multirow}
\usepackage{bm}
\usepackage{amsthm}

 \usepackage[most]{tcolorbox}

\newtcolorbox{myquotebox}{
  colback=white!0, 
  colframe=black, 
  rounded corners,
  boxrule=0.5pt, 
  title=Prompt:,
  left=2mm, 
  right=2mm, 
  top=1mm, 
  bottom=1mm 
}
\usepackage{todonotes}
\usepackage{xspace}

\definecolor{lightgrey}{RGB}{158, 158, 158}
\definecolor{goldenrod}{rgb}{0,0,0.8}
\definecolor{deepred}{rgb}{0.6,0,0}
\definecolor{deepgreen}{rgb}{0,0.5,0}
\definecolor{pink}{RGB}{219, 48, 122}
\definecolor{forestgreen}{RGB}{34,139,34}
\definecolor{goldenrod}{RGB}{218,165,32}
\definecolor{sepia}{RGB}{112,66,20}

\crefname{figure}{Fig.}{Figs.}
\crefname{table}{Table}{Tables}
\crefname{appendix}{App.}{App.}
\crefname{section}{§}{§§}
\crefname{equation}{Eq.}{Eqs.}

\newcommand\myparagraph[1]{
\vskip 0.05in 
\noindent{\bf {#1}}}
\newcommand*\samethanks[1][\value{footnote}]{\footnotemark[#1]}

\title{DIRAS: Efficient LLM Annotation of Document Relevance \\ for Retrieval Augmented Generation}

\author{ 
    Jingwei Ni\textsuperscript{\rm 1,2}\thanks{Equal Contributions.},
    Tobias Schimanski\textsuperscript{\rm 2}\samethanks, Meihong Lin\textsuperscript{\rm 4}, \\
    \textbf{Mrinmaya Sachan}\textsuperscript{\rm 1},
    \textbf{Elliott Ash}\textsuperscript{\rm 1},
    \textbf{Markus Leippold}\textsuperscript{\rm 2,3}  \\
    \textsuperscript{\rm 1}ETH Zurich \hspace{5mm}
    \textsuperscript{\rm 2}University of Zurich \hspace{5mm} 
    \textsuperscript{\rm 3} Swiss Finance Institute (SFI) \\
    \textsuperscript{\rm 4} University of
Electronic Science and Technology of China \\ 
    \texttt{\{jingni, msachan, ashe\}@ethz.ch, meihong\_lin@uestc.edu.ch}
    \\\texttt{\{tobias.schimanski, markus.leippold\}@df.uzh.ch}
}

\begin{document}
\maketitle
\begin{abstract}
Retrieval Augmented Generation (RAG) is widely employed to ground responses to queries on domain-specific documents. But do RAG implementations leave out important information when answering queries that need an integrated analysis of information (e.g., \textit{Tell me good news in the stock market today.})? 
To address these concerns, RAG developers need to annotate information retrieval (IR) 
data for their domain of interest, which is challenging because (1) domain-specific queries usually need nuanced definitions of relevance beyond shallow semantic relevance; and (2) human or GPT-4 annotation is costly and cannot cover all (query, document) pairs (i.e., annotation selection bias), thus harming the effectiveness in evaluating IR recall. 
To address these challenges, we propose DIRAS (\textbf{D}omain-specific \textbf{I}nformation \textbf{R}etrieval \textbf{A}nnotation with \textbf{S}calability), a manual-annotation-free schema that fine-tunes open-sourced LLMs to consider nuanced relevance definition and annotate (partial) relevance labels with calibrated relevance scores. Extensive evaluation shows that DIRAS enables smaller (8B) LLMs to achieve GPT-4-level performance on annotating and ranking unseen (query, document) pairs, and is helpful for real-world RAG development.
\ifarxiv
\footnote{All code, LLM generations, and human annotations in \url{https://github.com/EdisonNi-hku/DIRAS}.}
\else
\footnote{All code, LLM generations, and human annotations in \url{https://anonymous.4open.science/r/DIRAS-6D49}.}
\fi
\end{abstract}

\section{Introduction} \label{sec:introduction}

RAG has become one of the most popular paradigms for NLP applications \citep{gaoRetrievalAugmentedGenerationLarge2024}. One core phase of RAG systems is Information Retrieval (IR), which leverages cheap retrievers to filter relevant information and thus save LLM inference costs. 
However, IR can be a performance bottleneck for RAG \citep{chenDenseRetrievalWhat2023,gaoRetrievalAugmentedGenerationLarge2024}. Both leaving out important relevant information (\textit{low recall}) as well as including excessively related but irrelevant information (\textit{low precision}) may lead to severe decrease in performance \citep{niCHATREPORTDemocratizingSustainability2023,cuconasuPowerNoiseRedefining2024,niu2024ragtruth, schimanski2024faithful}. Furthermore, evaluation results on general-domain benchmarks \citep{thakurBEIRHeterogenousBenchmark2021} may hardly indicate the IR performance on RAG systems, as the definition of relevance varies drastically across different domains and use cases \citep{schimanski2024climretrieve, relevance_definition}. To address these concerns, \citet{saadfalcon2023ares} propose ARES to fine-tune an in-domain LM judge to evaluate context relevance. Although showing effectiveness in evaluating RAG systems for QA datasets like HotpotQA \citep{yang2018hotpotqadatasetdiverseexplainable}, ARES does not address two significant \textbf{C}hallenges that real-world RAGs are faced with:

\begin{figure}[t]
    \centering	\includegraphics[width=\columnwidth]{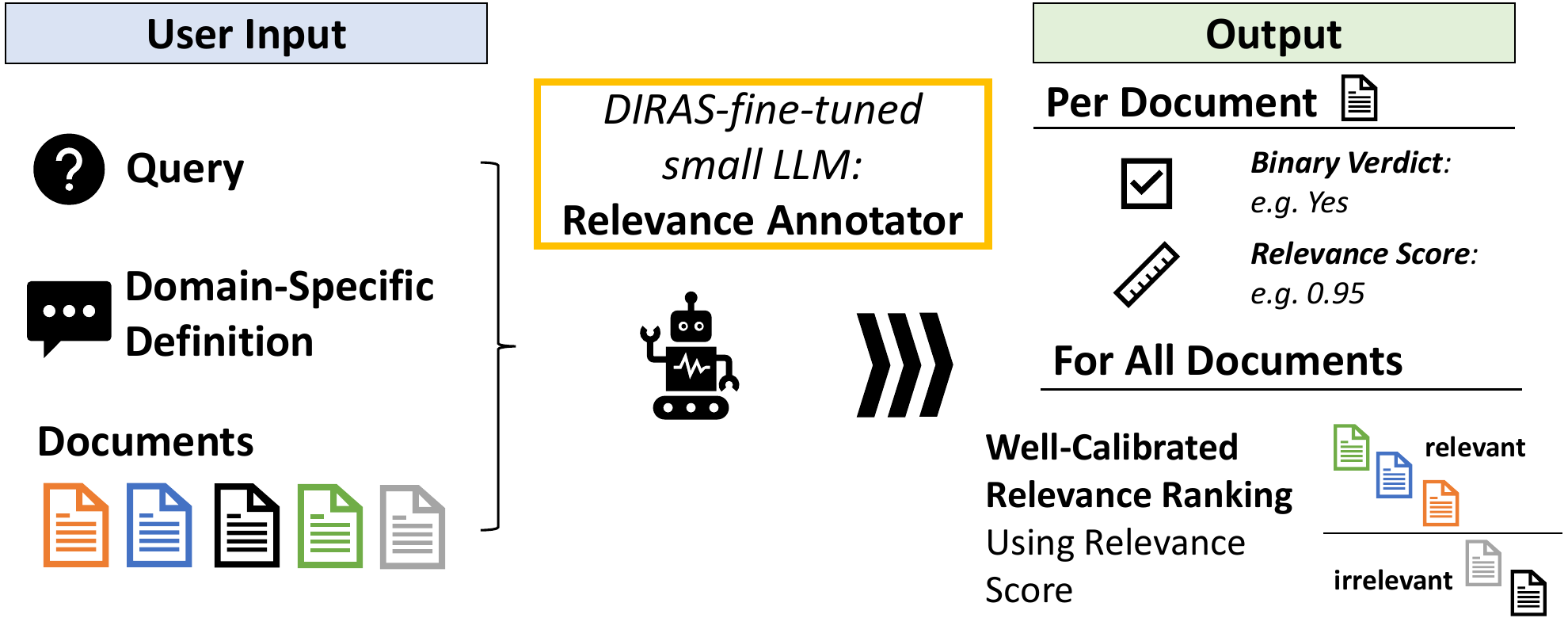}
	\vspace{-1em}
    \caption{ Overview of the functionality of DIRAS taking (query, relevance definition, document) triplets as input and output a binary verdict and a well-calibrated relevance score, which is sensitive to the grey-scale of partial relevance.}
	\label{fig:func_DIRAS}
\end{figure}

\myparagraph{C1. IR Recall}: RAG systems are more generally purposed than QA models, where many user queries require an integrative information analysis. For example, \textit{Write an overview of major events in World War 2}; or \textit{What is good news for the stock market today?} For such integrative queries, a good IR recall is necessary for comprehensive responses. \citeposs{saadfalcon2023ares} approach evaluates the relevance of retrieved contexts (precision), but is agnostic to other important information that the RAG retriever might leave out (recall). Besides, current RAG literature \citep{yan2024correctiveretrievalaugmentedgeneration,wang2024speculativeragenhancingretrieval} mostly relies on QA datasets \citep{joshi2017triviaqalargescaledistantly,yang2018hotpotqadatasetdiverseexplainable,dinan2019wizardwikipediaknowledgepoweredconversational,trivedi-etal-2022-musique} for evaluation, where the questions are less integrative and can mostly be answered by specific facts from one or few sources. For such questions, IR precision is more important than recall. 
As a result, the challenges of IR recall for integrative queries are heavily under-explored.

\myparagraph{C2. Relevance Definitions and Partial Relevance}: To thoroughly gather relevant information for integrative queries, the IR model should go beyond shallow semantic relationships and consider domain-specific relevance definitions. Furthermore, 
domain-specific requirements and subjectivity in IR annotation create a rich gray scale of \textit{partial relevance} between \textit{relevant} and \textit{irrelevant} (\citealp{relevance_definition,Saracevic2008EffectsOI,thomasLargeLanguageModels2024}; also see \cref{appendix:partial_relevance}). However, partial relevance is neglected entirely in RAG context relevance evaluation \citep{saadfalcon2023ares,es-etal-2024-ragas}. 

As a combined solution for these challenges, we propose \textbf{DIRAS}, a framework for efficient and effective relevance annotation. To address \textbf{C1}, DIRAS distills relevance annotation ability from SOTA generic teacher LLMs to small student LLMs, which can cost-efficiently annotate broad (query, document) pairs for IR recall evaluation. For better efficiency, student LLMs conduct pointwise annotation -- annotating (query, document) pairs one-by-one -- which is under-explored in related work \citep{sun2023instruction,qin2024large} but achieves good performance for relevance annotation. To address \textbf{C2}, DIRAS student LLMs are trained to comprehend nuanced relevance definitions, thus handling queries with various requirements. The student LLMs annotate binary relevance labels with well-calibrated relevance scores. Thus, the relevance scores can be used for relevance ranking and to calibrate the annotation accuracy \citep{ni2024afacta}. Thereby, these continuous relevance scores also address the grayscale of partial relevance (see \cref{fig:func_DIRAS} for an illustration of DIRAS functionality).

We evaluate DIRAS in three steps. First, we annotate ChatReportRetrieve to evaluate the design decisions for making DIRAS models optimized relevance annotators (\cref{sec:chatreport}). 
The evaluation shows that the fine-tuned student ($\leq$ 8B) LLMs effectively understand nuanced relevance definitions -- achieving GPT-4-level performance (\cref{sec:open_sourced_ft}). Second, we showcase how DIRAS assists in real-life IR annotation by re-annotating ClimRetrieve \citep{schimanski2024climretrieve}. Results show that DIRAS student LLMs can effectively capture partial relevance, leverage improved relevance definitions, mitigate annotation selection bias, and annotate benchmarking datasets for IR algorithms (\cref{sec:exs_on_climretrieve}). Third, we re-annotate document relevance for general QA and RAG datasets, showing DIRAS's potential in generic domains (\cref{sec:qa_experiments}). Collectively, our contributions include:

\begin{enumerate}[itemsep=0pt,topsep=1pt]
\item We propose DIRAS, a framework fine-tuning open-sourced LLMs into efficient and effective IR annotators, taking domain expertise into account.
\item We annotate ChatReportRetrieve, the first IR benchmark addressing integrative queries in RAG, and providing explicit relevance definition as annotation guidelines.
\item We showcase how to apply DIRAS in real-world IR annotation by re-annotating ClimRetrieve and general QA datasets.
\end{enumerate}

\begin{figure*}[t]
    \centering
	\includegraphics[width=1\textwidth]{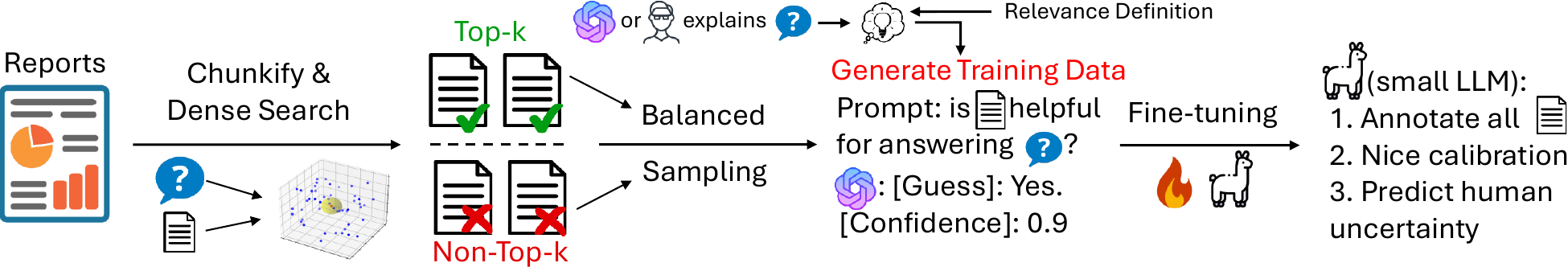}
	\caption{DIRAS pipeline. Domain-specific queries, and documents as inputs; calibrated student LLM annotators as outputs.}
	\label{fig:overview}
 \vspace{-0.7em}
\end{figure*}


\section{DIRAS Pipeline} 
\label{sec:pipeline}

The DIRAS pipeline is illustrated in \cref{fig:overview}. It comprises three steps: sampling a subset of (query, document) pairs for training data creation, obtaining relevance definitions for queries, and distilling relevance annotation ability from teacher to student LLMs. 

\myparagraph{Sampling (Query, Document) Pairs}: DIRAS takes domain-specific queries and documents as input. To sample representative (query, document) pairs as training data, it first ranks documents for each query using a small dense retriever. Then, it samples an equal number of documents within and outside of top-k (a pre-defined hyperparameter) to obtain representative documents for each query. While sampling in top-k aims at covering some relevant documents, sampling outside of top-k ensures covering the broader distribution of (query, document) pairs.

\myparagraph{Obtaining Relevance Definitions}: Each query in the sampled (query, document) pairs needs to be accompanied by an explicit definition indicating what is relevant or irrelevant to the question. The relevance definition can be generated by human experts, LLMs, or in a collaboration of both. In our experiments, we find GPT-4 generates suitable relevance definitions using the prompt in \cref{appendix:QuestionBackground}.

\myparagraph{Distilling Relevance Annotations from Teacher to Student}: With the sampled (query, definition, document) triplets, DIRAS creates relevance-annotation data with a SOTA generic teacher LLM $\mathcal{M}_{t}$ and the prompt template $\mathcal{P}$ (illustrated in \cref{fig:prompt_illustration}). 
Finally, the created data is used to fine-tune student LLMs $\mathcal{M}_{s}$, which will be used to conduct broad binary relevance annotation with confidence scores for calibration \citep{tianJustAskCalibration2023}. 

%

\section{Optimizing Relevance Annotation}
\label{sec:ex_chatreport}

To train a highly performant student LLM $\mathcal{M}_{s}$ for relevance annotation, we proceed in three steps. First, we annotate a novel task-specific dataset for our evaluation, ChatReportRetrieve (\cref{sec:chatreport}). Second, we assess design choices and related work baselines to find the best-performing strategy for our relevance-annotation data creation with the teacher model $\mathcal{M}_{t}$ (\cref{sec:prompt_selection}). Third, we optimize the fine-tuning of student LLMs $\mathcal{M}_{s}$ by investigating implementation variants (\cref{sec:open_sourced_ft}).

\subsection{ChatReportRetrieve} \label{sec:chatreport}

To evaluate the teacher LLMs' ($\mathcal{M}_{t}$) and student LLMs' ($\mathcal{M}_{s}$) comprehension of nuanced relevance definitions, we need to provide them with the same annotation guidelines (i.e., relevance definitions) and compare their annotation performance. 
To the best of our knowledge, there is no existing IR dataset that provides a nuanced relevance definition for each query. Hence, we annotate ChatReportRetrieve for our evaluation.

\myparagraph{Data and annotation guideline preparation}: ChatReportRetrieve is based on real-user integrative queries from ChatReport -- a chat tool for answering climate-related questions based on corporate reports\footnote{\url{https://reports.chatclimate.ai/}} \citep{niCHATREPORTDemocratizingSustainability2023}. We sample a wide range of climate reports and representative user queries about the reports to construct ChatReportRetrieve. Then, we conduct a train-test split, making sure no test set report or query is seen in the training set. Finally, we prompt GPT-4 to draft relevance definitions for all queries. GPT-4 drafted relevance definitions show a good understanding of the climate disclosure domain, according to a domain expert's feedback. See Appendices for details of data preprocessing (\cref{appendix:document_parsing}) and relevance definition generation (\cref{appendix:QuestionBackground}).

\myparagraph{Test Data Annotation}: We leverage relevance definitions as the annotation guidelines. 
\textit{If and only if} a document addresses the relevance definition, it is deemed as (partially) relevant. We explicitly account for partial relevance when the document addresses the periphery of the definition.
The data labeling process follows two steps. First, we employ two annotators who independently annotate all test data to be either relevant, irrelevant, or partially relevant. Second, we employ a subject-matter expert in corporate climate disclosure to resolve conflicts to obtain final relevance labels. Besides \textit{relevance labels}, we also obtain \textit{uncertainty labels} from human annotations: Whenever there is strong disagreement (co-existence of relevance and irrelevance labels) or agreement on partial relevance (two or more annotators agree on partial relevance), the data point is labeled as uncertain. Inter-annotator agreement and other details can be found in \cref{appendix:expert_annotation}.

\myparagraph{Evaluation Metrics}: LLM predictions contain a binary relevance annotation and a confidence score. They will be evaluated against \textit{relevance} or \textit{uncertainty} of ChatReportRetrieve labels on four dimensions: (1) \textbf{Binary Relevance}: We compute the F1 Score of models' binary relevance prediction using \textit{relevance labels}. Binary relevance labels are important for deciding which documents should be passed to LLMs. (2) \textbf{Calibration}: Confidence scores should calibrate the binary accuracy to indicate annotation quality. We use Expected Calibration Error (ECE), Brier Score, and AUROC to measure calibration performance, following \citet{kadavathLanguageModelsMostly2022} and \citet{tianJustAskCalibration2023}. (3) \textbf{Information Retrieval}: The confidence scores also give a calibrated relevance probability which can be used to rank documents for each query. To directly evaluate the ranking performance, we measure nDCG and MAP upon \textit{relevance labels}. (4) \textbf{Uncertainty}: If the models understand the difficulty caused by partial relevance, they should have lower confidence scores on samples that humans found uncertain. Thus we compute average precision (AP) scores between confidence and \textit{uncertainty labels}. Details of computing all metrics are in \cref{appendix:metrics_details}.

\begin{figure}[t]
\centering
\begin{tikzpicture}
\node[anchor=north west, draw, minimum width=\columnwidth, fill=white] (box) at (0,0) {
\begin{minipage}{0.95\columnwidth}
\small
\textbf{Prompt:} \\
\hangindent=0.5em \hangafter=1 \textcolor{blue}{<question>}: What is the firm's Scope 3 emission?\\
\hangindent=0.5em \hangafter=1 \textcolor{deepgreen}{<question\_definition>}: This question is looking for information about the firm's emission in ...\\
\hangindent=0.5em \hangafter=1 \textcolor{brown}{<paragraph>}: \{one text chunk from a climate report\}\\
\hangindent=0.5em \hangafter=1 Is \textcolor{brown}{<paragraph>} helpful for answering \textcolor{blue}{<question>}? Provide your best guess, and confidence score from 0 to 1.\vspace{0.6em}

\textbf{Teacher LLM $\mathcal{M}_{t}$:} \\
\hangindent=0.5em \hangafter=1 \textcolor{lightgrey}{[Reason]: \{Reason why the paragraph is (un)helpful.\}} \\
\hangindent=0.5em \hangafter=1 \textcolor{deepred}{[Guess]: \{Yes or No.\}} \\
\hangindent=0.5em \hangafter=1 \textcolor{deepred}{[Confidence]: \{confidence score between 0.0 and 1.0.\}} 
\end{minipage}
};
\end{tikzpicture}
\caption{ Our prompt template $\mathcal{P}$ for distilling training data from $\mathcal{M}_{t}$. ``[Reason]'' is only used in the CoT setup. It is shortened for presentation. Full $\mathcal{P}$ is in App. \cref{fig:diras_prompt}.}
\label{fig:prompt_illustration}
\end{figure}

\subsection{Optimizing the Training Data Creation}
\label{sec:prompt_selection}
\begin{table}[t]
\small
\centering
\resizebox{\columnwidth}{!}{
\begin{tabular}{lccccc}
\toprule
\textbf{Setting}        & \textbf{Unc.} & \textbf{Bin.}  & \textbf{Cal.} & \textbf{Info.}    &\textbf{Avg.}            \\ \midrule
List-2/1  & - & - & - & 76.86 & -    \\
List-2/1-D  & - & - & - & 74.72 & -    \\ 
List-10/5  & - & - & - & 84.74 & -    \\
List-10/5-D  & - & - & - & 84.45 & -    \\ 
List-20/10  & - & - & - & 78.05 & -    \\
List-20/10-D  & - & - & - & 82.54 & -    \\ \midrule
RAGAs & - & 68.15 & - & 37.13 & - \\
ARES-0-Shot & \underline{25.48} & \underline{52.63} & \underline{79.35} & \underline{77.67} & \underline{58.79} \\
ARES-2-Shot & 17.38 & 3.85 & 63.49 & 44.97 & 32.42 \\
ARES-4-Shot & 18.16 & 5.13 & 69.51 & 49.68 & 35.62 \\
ARES-8-Shot & 16.99 & 28.81 & 66.77 & 48.75 & 40.33 \\ 
ARES-16-Shot & 20.65 & 24.20 & 64.76 & 44.85 & 38.61 \\\midrule
Point-Ask & 39.27 & 84.07 & 90.59 & 87.57 & 75.37 \\ 
Point-Ask-Prob-D & 44.74 & 84.52 & \textbf{91.31} & 88.39 & 77.24  \\ 
Point-Tok-D & 28.83 & \textcolor{gray}{86.32} & 84.48 & 80.90 & 70.53 \\
Point-Ask-D & \textbf{54.01} & \textbf{86.32} & \underline{91.10} & \textbf{88.48} & \textbf{80.00}  \\
\bottomrule
\end{tabular}
}
\caption{\label{table:prompt_selection} GPT-4's performance on ChatReportRetrieve test set with different ranking methods (Point- or Listwise), RAGAs and ARES-few-shot, with/without relevance definition (D), and calibration method (Ask or Tok). Bin., Cal., Info., and Unc. stand for evaluation dimensions in \cref{sec:chatreport}.
}
\vspace{-0.7em}
\end{table}

We aim to train a highly performant student LLM $\mathcal{M}_{s}$ for relevance annotation. Thus, it is crucial to identify best-performing implementation choices that can be used to distill high-quality training data from the teacher LLM $\mathcal{M}_{t}$. Specifically, we compare the following four implementation choices:


\myparagraph{ARES few-shot vs. relevance definitions}: ARES \citep{saadfalcon2023ares} and DIRAS have different strategies to create target-domain training data: the former uses few-shot ICL\footnote{Original ARES uses few-shot ICL to create synthetic queries instead of relevance labels, which is not applicable for ChatReportRetrieve. Thus we use their ChatGPT prompt for relevance judgement, while replacing GPT-3.5 with GPT-4.} while the latter uses relevance definitions. We also include RAGAs \citep{es-etal-2024-ragas} relevance judgement as a baseline.

\myparagraph{Pointwise vs. Listwise}: The listwise method is popular in ranking data creation given its moderate cost and good performance \citep{sun-etal-2023-chatgpt,pradeep2023rankzephyr}. However, the more efficient pointwise method is under-explored in prior work -- majorly due to the concern about poor calibration \citep{sun2023instruction,qin2024large}.

\myparagraph{Calibration method (Tok vs. Ask)}: One calibration method is to get the relevance confidence by probing the model's generation probability of the token Yes/No when predicting a document's relevance (Tok, \citealp{liang2023holistic}). An alternative way is directly asking LLMs to verbalize confidence score, which may work better for instruction following LLMs (Ask, \citealp{tianJustAskCalibration2023}). 

\myparagraph{With vs. without relevance definition}: As ChatReportRetrieve test data is annotated based on the relevance definition, performance should increase if the model correctly takes the in-context relevance definition into consideration.

Following the takeaways of \citet{thomasLargeLanguageModels2024}, we design the prompt $\mathcal{P}$ for the pointwise method, relevance definition and CoT prompting (see \cref{fig:prompt_illustration} and \cref{fig:diras_prompt}, prompt without definition in \cref{fig:diras_prompt_no_e}). We use the listwise ranking prompt from \citet{sun-etal-2023-chatgpt} and \citet{pradeep2023rankzephyr} (see prompt with/without definition in \cref{fig:list_definition}/\cref{fig:list_original}). For the pointwise method, we run one variation to test prompt sensitivity: directly asking for relevance probability instead of confidence for guess (prompt in \cref{fig:diras_prompt_prob}). As the listwise ranking is sensitive to window/step size, we run three variations with window/step sizes of 2/1, 10/5, and 20/10. text-embedding-3-small is used for listwise methods' initial ranking. ARES and RAGAs settings are from the original papers. For few-shot ICL, we keep relevant/irrelevant samples balanced to avoid bias. 

\myparagraph{Takeaways}: Results in \cref{table:prompt_selection} show that: (1) Few-shot ICL fails to teach domain-specific relevance. The ICL illustrations (even balanced) seem to bias GPT-4 to underperform the zero-shot setting. (2) With the proper calibration method (Ask), the pointwise method outperforms the listwise method. (3) The listwise method is sensitive to window size, while the pointwise method gives more consistent performance across prompts. (4) Adding a relevance definition drops the listwise performance in 2 out of 3 cases, while that improves the pointwise performance. Thus we choose pointwise to be our distillation strategy.

\subsection{Optimizing DIRAS student LLMs}
\label{sec:open_sourced_ft}

\begin{table}[t]
\small
\centering
\resizebox{\columnwidth}{!}{
\begin{tabular}{lccccc}
\toprule
\textbf{Setting}        & \textbf{Unc.} & \textbf{Bin.}  & \textbf{Cal.} & \textbf{Info.}    &\textbf{Avg.}            \\ \midrule
Small-embed & - & - & - & 66.34 & -    \\
Large-embed & - & - & - & 69.36 & -    \\
BGE-Gemma & - & - & - & 68.47 & -    \\
GPT-3.5  & 29.71 & 45.27 & 85.46 & 74.16 & 58.65    \\
GPT-4  & \textbf{54.01} & \textbf{86.32} & \underline{91.10} & \underline{88.48} & \textbf{80.00}    \\ \midrule
Llama3-CoT-Ask  & 36.57 & 76.58 & 89.30 & 86.15 & 72.15    \\ 
Llama3-CoT-Tok  & 41.74 & 76.58 & 86.61 & 85.96 & 72.72    \\ 
Llama3-Ask  & 40.18 & \underline{82.11} & 90.14 & 86.02 & 74.61     \\ 
Llama3-Tok  & 41.60 & \underline{82.11} & \textbf{91.35} & \textbf{89.19} & \underline{76.06}$^\dagger$    \\ \midrule
Phi3-CoT-Ask  & 36.08 & 72.95 & 88.76 & 80.56 & 69.59    \\ 
Phi3-CoT-Tok  & 35.49 & 72.95 & 84.20 & 80.64 & 68.32    \\ 
Phi3-Ask  & 32.30 & 73.23 & 85.56 & 80.05 & 67.79    \\ 
Phi3-Tok  & 38.00 & 73.23 & 89.52 & 86.94 & 71.92$^\dagger$    \\ \midrule
Gemma-CoT-Ask  & 31.60 & 72.38 & 86.38 & 81.39 & 67.94    \\ 
Gemma-CoT-Tok  & 39.03 & 72.38 & 83.49 & 80.33 & 68.81    \\ 
Gemma-Ask  & 25.74 & 67.13 & 81.80 & 77.43 & 63.02    \\ 
Gemma-Tok  & \underline{50.72} & 67.13 & 90.07 & 81.17 & 72.27$^\dagger$    \\ 
\bottomrule
\end{tabular}
}
\caption{\label{table:main_results} Comparison between the fine-tuned student $\mathcal{M}_{s}$ and different baselines on ChatReportRetrieve test data. 
The best scores are \textbf{bolded} and the second bests are \underline{underlined}.$\dagger$ denotes the best score achieved by each backbone LLM.}
\vspace{-0.7em}
\end{table}

DIRAS student LLMs $\mathcal{M}_{s}$ will be used to annotate all (query, document) combinations. 
Two methodological choices might influence the quality of fine-tuned student LLMs. First, we explore the role of Chain-of-Thought (CoT) reasoning.
Second, we investigate the choice of calibration method \citep{tianJustAskCalibration2023}. To explore these aspects, we fine-tune $\mathcal{M}_{s}$ in four settings: $\mathcal{M}_{s}$-CoT-Ask, $\mathcal{M}_{s}$-CoT-Tok, $\mathcal{M}_{s}$-Ask, $\mathcal{M}_{s}$-Tok, where CoT means $\mathcal{M}_{s}$ is tuned to generate [Reason], [Guess], and [Confidence]; without CoT denotes $\mathcal{M}_{s}$ is tuned to only generate [Guess] and [Confidence]; ``Ask'' means the result is calibrated by the generated confidence score in [Confidence] field; and ``Tok'' means we take the token-level probability of ``Yes/No'' after ``[Guess]:'' as the confidence score for calibration. The prompt in \cref{fig:prompt_illustration} is used for fine-tuning. The ``[Reason]:'' line is removed in settings without CoT. 

We fine-tune Llama-3-8B-instruct \citep{llama3modelcard}, gemma-7b-it \citep{gemmateam2024gemma}, and Phi-3-mini-4k-instruct \citep{abdin2024phi3} (details in \cref{appendix:ft_detail}). We compare these fine-tuned student models with baselines including GPT-3.5 and GPT-4 using prompt $\mathcal{P}$; the OpenAI embedding models text-embedding-3-small, and text-embedding-3-large; and BGE Gemma reranker\footnote{https://huggingface.co/BAAI/bge-reranker-v2-gemma}, a popular LLM-based reranker for general domain. 

\begin{figure}[t]
    \centering
	\includegraphics[width=\columnwidth]{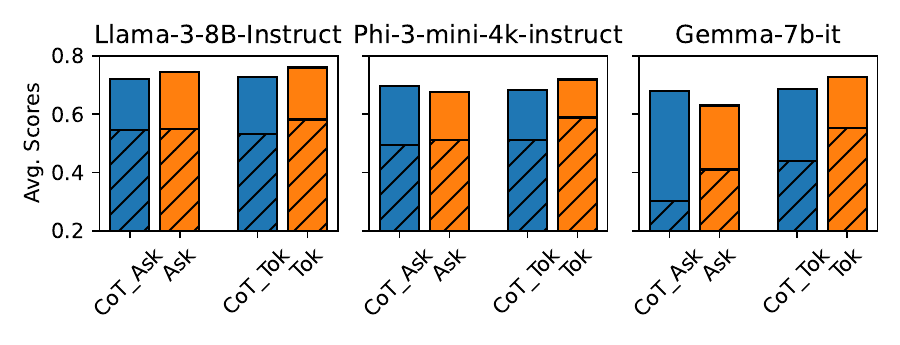}
 \vspace{-2em}
	\caption{Shaded bars denote the performance of original models. Colored bars denote the improvement brought by fine-tuning.}
	\label{fig:finetune_improvement}
 \vspace{-0.2em}
\end{figure}

As \cref{fig:finetune_improvement} shows, fine-tuning improves original models in all settings. Furthermore, \cref{table:main_results} shows the results of all fine-tuned student models in comparison to all baselines. We observe that $\mathcal{M}_{s}$-Tok outperforms other settings for all LLM architectures. The best setting Llama3-Tok achieves GPT-4 level performance in calibration and IR on unseen questions and reports.

Interestingly, we find that omitting the chain of thought usually leads to a performance increase for all three LLM architectures. CoT sometimes leads to a limited increase when asking for calibration (Ask), but constantly results in a performance drop when calibrated with token-level probability (Tok). Therefore, $\mathcal{M}_{s}$ should be fine-tuned without CoT for inference efficiency. Moreover, Tok rarely underperforms Ask, different from \citet{tianJustAskCalibration2023}'s finding and our observations in \cref{table:prompt_selection}. Thus, future work may consider probabilities of important tokens (e.g., Yes/No in our prompt template) as a promising calibration tool.

\section{Real-World Applications} \label{sec:application_real_world}
Having established and benchmarked the design choices of DIRAS student LLMs, we now investigate their real-world application capabilities. First, we showcase how the DIRAS $\mathcal{M}_{s}$ can assist IR annotation in a real-world setting, leveraging the ClimRetrieve dataset \citep{schimanski2024climretrieve} (\cref{sec:exs_on_climretrieve}). Second, we show the DIRAS pipeline is also applicable for general domain QA (\cref{sec:qa_experiments}).

\subsection{Applying to ClimRetrieve} \label{sec:exs_on_climretrieve}
ClimRetrieve \citep{schimanski2024climretrieve} records human analysts' real-life workflow of reading full reports, searching for relevant information, and annotating useful information for climate-related questions with relevance scores 1-3. ClimRetrieve contains 43K (query, document) pairs (8K unique documents but each document in a report is multiplied by the amount of analyzed queries per report) out of which 595 are gold labels for relevant (query, document) pairs. Other not annotated (query, document) combinations might be either irrelevant or a part of annotation selection bias -- a widely existing problem in IR annotation \citep{thakurBEIRHeterogenousBenchmark2021}. To succeed on this dataset, the IR model needs to (1) capture the analysts' mental model about useful information (i.e., relevance definition), 
and (2) understand fine-grained differences in degree of relevance (score 1-3).

Since ClimRetrieve is still in the climate domain, we re-annotate its data with the best $\mathcal{M}_{s}$ in \cref{sec:ex_chatreport} (Llama3-Tok), and explore whether $\mathcal{M}_{s}$'s annotations can (1) \textbf{RQ1}: reflect fine-grained differences in degree of relevance; (2) \textbf{RQ2}: be improved through refining relevance definitions; (3) \textbf{RQ3}: mitigate annotation selection bias for better IR recall evaluation; and (4) \textbf{RQ4}: benchmark and select IR algorithms.

\begin{table}[t]
\small
\centering
\resizebox{\columnwidth}{!}{
\begin{tabular}{lcccc}
\toprule
\textbf{Setting}        & \textbf{nDCG} & \textbf{nDCG@5}     &\textbf{nDCG@10} & \textbf{nDCG@15}\\ \midrule
Random  & 71.04
 &            50.88
 &             52.77
 &             54.45
 \\
Small-embed & 74.52
 &            61.28
 &             60.36
 &             61.69
 \\
Large-embed & 76.30
 &            63.13
 &             63.36
 &             64.67
 \\
GPT-3.5  &          74.62
 &            60.08
 &             61.49
 &             61.91
 \\
GPT-4  &          75.55
 &           60.89
 &             63.23
 &             65.26 \\
Llama3-Ask & \textbf{77.23} &    \textbf{67.60} &             \underline{66.18} &             \textbf{67.57} \\
Llama3-Tok & \underline{76.55} &            \underline{67.20} &             \textbf{66.23} &        \underline{65.83}
 \\
\bottomrule
\end{tabular}
}
\caption{\label{table:ClimRetrieve_test} Performance on ranking the \textbf{relevant} (query, document) pairs in ClimRetrieve.}
\vspace{-0.7em}
\end{table}

\myparagraph{RQ1: Reflecting Fine-Grained Relevance Levels}.
We first evaluate Llama3-Tok's annotation on 595 gold labels of ClimRetrieve to verify whether it can effectively recover analysts' ranking for relevant content by understanding which documents are more helpful than others. Relevance definitions are drafted with GPT-4 with the same procedure as \cref{sec:ex_chatreport}. We report nDCG\footnote{MAP can only measure binary relevance and since we only investigate relevant samples, it cannot be calculated.} scores to measure the ranking performance on ClimRetrieve. Gold labels 1, 2, and 3 are assigned with relevance scores 1/3, 2/3, and 1. Besides OpenAI 3rd generation embedding models, we also have a random baseline where all (query, document) pairs are assigned a random relevance score between 0 and 1. The random baseline results are averaged over 5 random seeds (40 to 44). Importantly, all ClimRetrieve annotations are to some degree relevant, so improvement over the random baseline is challenging as the system needs to understand the trivial different degrees of relevance.

\cref{table:ClimRetrieve_test} presents different systems' performance. There is a clear trend of outperformance of the fine-tuned Llama-3 models in this challenging setting. 

\begin{table}[]
\centering
\small
\begin{tabular}{lcc}
\toprule
\textbf{Setting} & \textbf{nDCG} & \textbf{MAP} \\
\midrule
Llama3-Ask$_{\textit{generic}}$ & 29.95 & 26.51 \\
Llama3-Ask$_{\textit{improved}}$ & \textbf{30.89} & \textbf{29.31} \\ \midrule
Llama3-Tok$_{\textit{generic}}$ & 31.17 & 28.73 \\
Llama3-Tok$_{\textit{improved}}$ & \textbf{32.53} & \textbf{32.65} \\
\bottomrule
\end{tabular}
\caption{Comparison of using the generic and the improved relevance definitions for ranking \textbf{all} ClimRetrieve (query, document) pairs.}
\label{tab:comp_all_gen_spec_short}
\vspace{-1.3em}
\end{table}

\myparagraph{RQ2: Improving Performance through Improving Definitions}. 
So far, we used GPT-4 to draft the relevance definitions. To investigate the effect of improved definitions, we compare two setups: (1) The \textit{generic} relevance definition: the definition drafted by GPT-4. (2) The \textit{improved} relevance definition: The only way to improve the definition is to align it closer to ClimRetrieve annotators' mental model of document relevance. We achieve this by adding relevant text samples to the prompt for generating the definition with GPT-4 (see \cref{appendix:specific_query_definition} for details).

After creating the improved definitions, we repeat predicting the relevance scores with Llama3-Tok. Since we involve examples with various relevance scores to improve relevance definitions, these definitions might not help distinguish the granular level of partial relevance. However, the improved definitions might especially help to distinguish relevant documents from irrelevant ones. Calculating the nDCG and MAP score for all 43K (query, document) pairs, we find evidence for this notion (see \cref{tab:comp_all_gen_spec_short}). Thus, the inclusion of improved definitions seems to enhance the performance (see \cref{appendix:map_nDCG_qeustion_definition} for details).

\myparagraph{RQ3: Mitigating Annotation Selection Bias}.
ClimRetrieve employs a real-world analyst scenario. This entails that the human only selectively annotates documents that are likely to be relevant and assumes unannotated documents as irrelevant \citep[see e.g.,][]{thakurBEIRHeterogenousBenchmark2021}. Therefore, the dataset allows us to investigate our model's capabilities to counteract biases. For this purpose, we sample 200 disagreements between Llama3-Tok's annotation and the original ClimRetrieve labels. Then, we reannotate these samples with a human labler. To account for different confidence levels in Llama3-Tok's prediction, we differentiate prediction with confidence higher or lower than 0.95.

As \cref{tab:climeretrieve_disagreement} indicates, the model can be successfully used to overturn decisions of unseen, as irrelevant assumed documents (91.30\% for confidence $> 0.95$). Strikingly, even samples annotated by humans, i.e., those labeled as relevant, can be overturned, though with a lower certainty. We attribute this to differences in the unknown mental model of the ClimRetrieve labeler and our explicit relevance definitions (for details, see \cref{appendix:error_analysis}). However, for us, it is reaffirming to observe that the DIRAS model is consistent with its own definition. Thus, we conclude that DIRAS' labeling is effective and helps to mitigate annotation selection bias.

\begin{table}[]
\centering
\small
\begin{tabular}{cccc}
\toprule
                       & \multicolumn{3}{c}{\textbf{ClimRetrieve (62.00\%)}} \\ \cline{2-4} \addlinespace[2pt] 
                       \vspace{-0.2em}
                       & \textbf{All}    & \textbf{Rel.}   & \textbf{Irr.}   \\ \midrule
Conf $\leq0.95$   & 36.84           & 27.90           & 48.48           \\
Conf $>0.95$ & \textbf{85.48}  & \textbf{78.18}  & \textbf{91.30}  \\ \bottomrule
\end{tabular}
\caption{Accuracy of student model Llama3-Tok annotations that disagree with original ClimRetrieve relevance labels. \textbf{All} denotes all sampled disagreed labels. \textbf{Rel.} ( \textbf{Irr.}) denotes the subset where the original label is \textit{relevant} (\textit{irrelevant}). Conf $\leq0.95$ ($>0.95$) denotes the subset where Llama3-Tok's confidence is lower (higher) than 0.95. \textbf{ClimRetrieve} (62.12\%) means that 62.12\% of data are annotated with $>0.95$ confidence.}
\label{tab:climeretrieve_disagreement}
\end{table}

\begin{table}[t]
\small
\centering
\begin{tabular}{lc}
\toprule
\textbf{Setting}        & \textbf{Kendall's} $\mathbf{\tau}$ \\ \midrule
BGE-Base  & 35.71  \\
BGE-Base-ft  & \textbf{36.34}  \\ \midrule
BGE-Large  & 34.74  \\
BGE-Large-ft  & \textbf{36.55}  \\ 
\bottomrule
\end{tabular}
\caption{\label{table:benchmark} Different embedding models' performance benchmarked by student model $\mathcal{M}_{s}$'s prediction on all 43K (query, document) pairs of ClimRetrieve. ``ft'' denotes the model is fine-tuned on in-domain data.}
\vspace{-0.7em}
\end{table}

\begin{table*}[ht]
\renewcommand{\arraystretch}{1} 
\small
\centering
\resizebox{\textwidth}{!}{
\begin{tabular}{lccccccccccccccc}
\toprule
& \multicolumn{3}{c}{\textbf{ELI5}} & & \multicolumn{3}{c}{\textbf{ASQA}} & & \multicolumn{3}{c}{\textbf{QAMPARI}} & & \multicolumn{3}{c} {\textbf{RAG-Bench}} \\
\cline{2-4} \cline{6-8} \cline{10-12} \cline{14-16} \addlinespace[2pt]  
\vspace{-0.2em}
& \textbf{N} & \textbf{N@5} & \textbf{N@10} & & \textbf{N} & \textbf{N@5} & \textbf{N@10} & & \textbf{N} & \textbf{N@5} & \textbf{N@10} & & \textbf{N} & \textbf{N@5} & \textbf{N@10} \\

\midrule
GPT-4 &  48.43 &   15.97 &    17.81 & & 64.62 &   38.82 &    46.32 & & 56.21 &      27.54 &       35.34 & & 42.91 &        31.84 &         41.35 \\
Llama3-Tok & \textbf{48.73} &   \textbf{17.04} &    \textbf{20.08} & & \textbf{64.90} &   \textbf{39.37} &    \textbf{48.44} & & \textbf{56.58} &      \textbf{28.70} &       \textbf{35.49} & & \textbf{48.27} &        \textbf{41.13} &        \textbf{47.88} \\

\bottomrule
\end{tabular}
}
\caption{Applying DIRAS to QA datasets, the IR performance of student model Llama3-Tok and GPT-4. \textbf{N} denotes nDCG.}
\label{tab:ir_on_qa}
\vspace{-0.7em}
\end{table*}

\begin{table*}[ht]
\renewcommand{\arraystretch}{1} 
\small
\centering
\resizebox{\textwidth}{!}{
\begin{tabular}{lccccccccccccccc}
\toprule
& \multicolumn{3}{c}{\textbf{ELI5} (85.05\%)} & & \multicolumn{3}{c}{\textbf{ASQA} (66.32\%)} & & \multicolumn{3}{c}{\textbf{QAMPARI} (72.66\%)} & & \multicolumn{3}{c} {\textbf{RAG-Bench} (60.33\%)} \\
\cline{2-4} \cline{6-8} \cline{10-12} \cline{14-16} \addlinespace[2pt]  
\vspace{-0.2em}
& \textbf{All} & \textbf{Rel.} & \textbf{Irr.} & & \textbf{All} & \textbf{Rel.} & \textbf{Irr.} & & \textbf{All} & \textbf{Rel.} & \textbf{Irr.} & & \textbf{All} & \textbf{Rel.} & \textbf{Irr.} \\

\midrule
Conf $\leq0.95$ & 62.35 &  48.48 &  71.15 & & 67.27 &  71.43 &   66.67 & & 74.11 &    73.33 &      74.23 & & 58.41 &    38.10 &         63.04 \\
Conf $>0.95$ & \textbf{84.71} &   \textbf{83.12} &    \textbf{100.0} & & \textbf{90.09} &   \textbf{95.83} &    \textbf{88.51} & & \textbf{90.00} &      \textbf{81.25} &       \textbf{91.49} & & \textbf{96.36} &        \textbf{95.24} &        \textbf{96.63} \\

\bottomrule
\end{tabular}
}
\caption{Accuracy of student model Llama3-Tok annotations that disagree with original relevance labels. Same setup as \cref{tab:climeretrieve_disagreement}.}
\label{tab:qa_disagreement}
\end{table*}

\myparagraph{RQ4: Benchmarking IR}. We use $\mathcal{M}_{s}$ to annotate all 43K ClimRetrieve datapoints and obtain a benchmarking dataset to select IR algorithms. This approach can be especially helpful when lacking human annotation and annotation selection bias is prevalent. Specifically, we compare the performance of embedding models before and after in-domain fine-tuning. If the $\mathcal{M}_{s}$-annotated benchmark gives higher scores to the fine-tuned checkpoints, that means it is capable of selecting a better model for this specific domain.

For this experiment, we first fine-tune open-sourced embedding models bge-large-en-v1.5 and bge-base-en-v1.5 \citep{chen2024bge} on ChatReportRetrieve test set\footnote{We fine-tune on the test instead of the training set to (1) leverage high-quality human annotation for fine-tuning; and (2) avoid indirect data leakage as $\mathcal{M}_{s}$ is fine-tuned on the training set.} (fine-tuning details in \cref{appendix:embedding_ft}). We then compare embedding models' relevance ranking with the predicted ranking of Llama3-Tok on all 43K (query, document) pairs in ClimRetrieve. We use Kendall's $\tau$ as the metric, which directly compares the correlation between two ranks. The results are shown in \cref{table:benchmark}. We find the Llama3-Tok-annotated benchmark successfully picks out the fine-tuned checkpoints, showing a capability of benchmarking information retrieval algorithms. 
\ifarxiv
Interestingly, the unfine-tuned BGE-Base correlates more to Llama3-Tok compared to BGE-Large, although the latter shows stronger performance on MTEB \citep{muennighoff-etal-2023-mteb}. This indicates the necessity of domain-specific benchmarking to tell the in-domain performance. 
\else

\fi

\subsection{Applying to QA Datasets} \label{sec:qa_experiments}

In this section, we apply the DIRAS pipeline to QA datasets that are widely used in RAG benchmarking. DIRAS addresses queries for broad information and IR recall. Thus, we include long-form QA datasets from ALCE \citep{gao2023enablinglargelanguagemodels}, including ELI5 \citep{fan2019eli5longformquestion}, ASQA \citep{stelmakh2023asqafactoidquestionsmeet}, and QAMPARI \citep{amouyal2023qampariopendomainquestionanswering}. We also include RAG-Bench \citep{fang2024enhancingnoiserobustnessretrievalaugmented} that consists of questions from TriviaQA \citep{joshi2017triviaqalargescaledistantly}, WebQ \citep{berant-etal-2013-semantic}, and Natural Questions \citep{kwiatkowski-etal-2019-natural}. RAG-Bench is chosen since it has labels for partial vs. full relevance, which is a focus of DIRAS. Importantly, the context relevance labels from ALCE and RAG-Bench are derived from reference answers to questions, using heuristics instead of manual annotation. Thus, they are to some extent \textbf{noisy}.

We first run the DIRAS pipeline\footnote{Different from Climate change datasets, QA datasets do not require nuanced relevance definition. Thus we use a fixed relevance definition: \textit{``the document is helpful only if its content answers the query''}. See full prompt in \cref{fig:qa_prompt}.} on each dataset to obtain corresponding Llama3-Tok models. Then we compare them with the teacher model -- GPT-4. Results in \cref{tab:ir_on_qa} show that Llama3-Tok outperforms GPT-4 in IR. Then for each dataset, we repeat the annotation selection bias assessment of RQ3 in \cref{sec:exs_on_climretrieve}. Again, we sample 200 disagreements between Llama3-Tok annotation and the original (potentially noisy) relevance labels, and manually check whether Llama3 or original labels are correct. As shown in \cref{tab:qa_disagreement}, Llama3-Tok's annotations are predominately correct with a confidence $>0.95$  (e.g., $85.05\%$ of ELI5). When there is a disagreement, relying on Llama3-Tok leads to less error (Acc. $>50\%$), especially when the confidence $>0.95$, thanks to the good calibration. For ASQA, QAMPARI, and RAG-Bench, the majority ($>90\%$) of the disagreement lies in originally irrelevant labeled part of the dataset (\textbf{Irr.}), possibly due to (query, document) pairs are selectively annotated. DIRAS achieves high accuracy in \textbf{Irr.} disagreements. Therefore, we reaffirm the notion that applying DIRAS to annotate broader (query, document) pairs can effectively reduce annotation selection bias, and thus improve IR recall benchmarking.
All implementation details are in \cref{appendix:details_qa}.

\ifarxiv
\section{Recommendation for Future RAG} \label{sec:recommondation}
\myparagraph{Avoiding Top-K Retrieval}: Naive RAG systems \citep{niCHATREPORTDemocratizingSustainability2023} usually retrieve top-k (a fixed number k) documents to augment LLM generation. However, different questions tend to have different amounts of relevant information.
Advanced RAG employs query routers to pick retrieval strategies \citep{gaoRetrievalAugmentedGenerationLarge2024}. However, choosing the proper k without access to full documents is still hard. To demonstrate this, we average the relevance score (predicted by Llama3-Tok) over all documents for each question in ClimRetrieve. The resulting average relevance score will be a proxy for the amount of relevant information on the question in all reports. As  \cref{fig:amt_relevant_info} shows, different questions vary considerably in the amount of relevant information. Therefore, we suggest not using top-k IR, avoiding the prior determined k that does not fit the actual amount of relevant information.

Given the calibrated prediction of DIRAS $\mathcal{M}_{s}$, an alternative way is to retrieve all documents whose relevance scores exceed a pre-defined threshold. Thus, different questions can retrieve different amounts of information depending on whether passing the relevance threshold. Advanced RAG designs can even strategically pick the calibrated threshold for different questions, for example, allowing more partial relevance for summary queries. \cref{fig:f1_threshold} shows the F1 Scores of GPT-4 and Llama3-Tok with different relevance thresholds. Llama3-Tok achieves good F1 scores over a wide range of thresholds. Thanks to its compact size (8B), it can be efficiently deployed as a reranker in RAG systems.

\myparagraph{Optimizing Relevance Definitions}: Results in \cref{table:main_results} and \cref{table:ClimRetrieve_test} are obtained with GPT-4-drafted relevance definitions (i.e., relevance definitions). Although this approach is useful in large-scale applications, there is still space for improvement by optimizing relevance definition, as shown in \cref{sec:exs_on_climretrieve}. According to \citet{relevance_definition}, the question originators are the gold standard for relevance definition. Hence, with the help of DIRAS, future RAG systems may allow users to customize their requirements for relevant information.

\begin{figure}[t]
    \centering
	\includegraphics[width=\columnwidth]{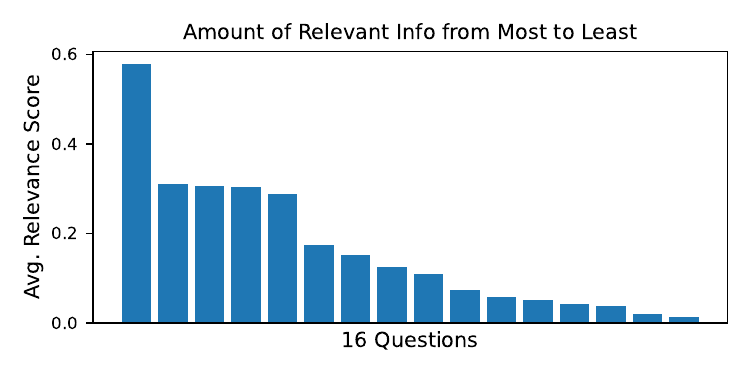}
 \vspace{-2em}
	\caption{The proximate amount of relevant information for 16 questions in all ClimRetrieve reports, according to Llama3-Tok's relevance scores.}
	\label{fig:amt_relevant_info}
 \vspace{-0.8em}
\end{figure}

\begin{figure}[t]
    \centering
	\includegraphics[width=\columnwidth]{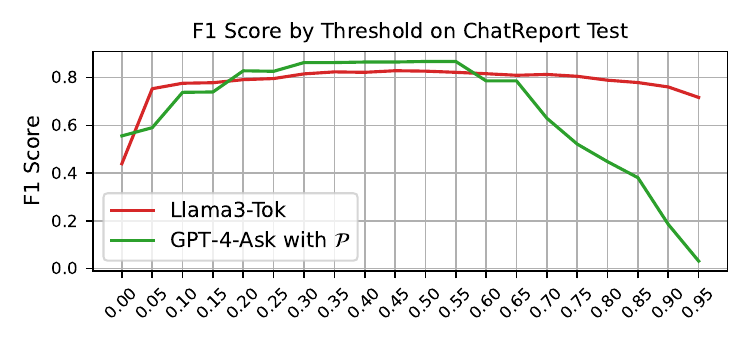}
 \vspace{-2em}
	\caption{Instead of always retrieving top-k, we can retrieve documents if they have relevance scores higher than a threshold. This figure shows the change of F1 scores for obtaining relevant documents by thresholds.}
	\label{fig:f1_threshold}
 \vspace{-1em}
\end{figure}
\else

\fi

\section{Background and Related Work}
\ifarxiv
IR plays an important role in RAG but also becomes a performance bottleneck \citep{gaoRetrievalAugmentedGenerationLarge2024}. Low precision in IR may cause LLMs to hallucinate or pick up irrelevant information \citep{cuconasuPowerNoiseRedefining2024,schimanski2024faithful}. Low recall may leave out critical information for analysis \citep{niCHATREPORTDemocratizingSustainability2023}. Domain-specific knowledge is also important for retrieval performance \citep{tang2024needdomainspecificembeddingmodels}.
\else

\fi

Prior work on IR in RAG has already explored the idea of using LLMs to judge relevance. The closest to our work are RAGAs and ARES. \citet{es-etal-2024-ragas} develop RAGAs to evaluate the relevance of each sentence in a paragraph with a closed-source LLM and create an aggregated score by dividing the number of relevant sentences over all sentences. \citet{saad-falconARESAutomatedEvaluation2023} aim to bring a relevance judge to a target domain through few-shot in-context learning (ICL): they first generate synthetic questions to given target-domain passages, and then fine-tune small classifiers for relevance judgment. 
However, many real-world challenges remain unaddressed. Specifically, IR recall, partial relevance, and domain-specific relevance definitions are neglected. ARES appears to have target-domain IR evaluation, but the synthetic data approach focuses on less integrative queries: each question is generated given a single passage and hard negatives are passages sampled from the same document, which could be (partially) relevant for integrative queries asking for broader information. Furthermore, we also find that few-shot ICL fails to teach domain-specific relevance to LLMs (\cref{sec:prompt_selection}). 

Besides, \citet{sun-etal-2023-chatgpt,sun2023instruction,pradeep2023rankzephyr,qin2024large} find that SOTA generic LLMs are good rerankers and such ability can be distilled to open-sourced LLMs. These studies all focus on pairwise or listwise ranking methods, and discusses that pointwise methods may not work due to bad calibration \citep{qin2024large}. However, when it comes to relevance annotation instead of reranking, pointwise relevance prediction is more suitable because it: (1) analyzes and annotates (query, document) pairs one-by-one, thus is more efficient and may better consider relevance definitions; and (2) can annotate both document rank and binary relevance labels (relevant or irrelevant) which are important for RAG in order to decide what information should be passed to the generator. We also show pointwise annotation works better with proper calibration method \citep{tianJustAskCalibration2023}. 
\section{Conclusion}
In this work, we introduce the DIRAS pipeline to fine-tune open-source LLMs to calibrated annotators. The DIRAS approach has two significant advantages: (1) it is case-specialised allowing the incorporation of domain-specific knowledge into definitions, and (2) it helps to efficiently label a huge amount of documents with calibrated relevance scores.

\section*{Limitations}
As with every work, this has limitations. 
First, our results show that DIRAS fine-tuning grants small student LLMs GPT-4-level performance on specific domains, but GPT-4 is not guaranteed to be perfect in all domains and cases. In certain niche domains, it might be necessary to augment GPT-4 with domain knowledge or agentic designs to achieve human-level performance in relevance annotation, and then create reliable training data for DIRAS. Although performance not always guaranteed, LLM annotation for document relevance is still necessary due to the sheer volume of (query, document) pairs and selection bias of human annotation.

Second, this project focuses on text documents. This means we do not evaluate the performance of the DIRAS pipeline on graph and table content. While this also presents a general limitation of modern-day RAG systems, we believe it is a crucial future step to generalize DIRAS's idea of scalable information retrieval benchmarking to multi-modality. 

Our third limitation, and also a viable option to address multi-modality, lies in the recent introduction of long-context LLMs. These may make the role of information retrieval in RAG less crucial as entire documents can be used to answer a question. At the same time, we observe that long-context models are good in needle-in-a-haystack problems but not as good when multiplied needles exist \citep{geminiteam2024gemini}. Thus, even for long-context LLMs, an efficient system like DIRAS could enable improving algorithms for finding and using multiple relevant pieces of information or help improve the model's ability to do so.

\section*{Ethics Statement}
\myparagraph{Human Annotation}: In this work, all human annotators are Graduate, Doctorate researchers, or Professors who have good knowledge about scientific communication and entailment. They are officially hired and have full knowledge of the context and utility of the collected data. We adhered strictly to ethical guidelines, respecting the dignity, rights, safety, and well-being of all participants. 

\myparagraph{Data Privacy or Bias}: There are no data privacy issues or biases against certain demographics with regard to the data collected from real-world applications and LLM generations. All artifacts we use are under a creative commons license. We also notice no ethical risks associated with this work

\myparagraph{Reproducibility Statement}: To ensure full reproducibility, we will disclose all codes and data used in this project, as well as the LLM generations, GPT-4 and human annotations. For OpenAI models, we use ``gpt-4-0125-preview'' and ``gpt-3.5-turbo-0125''. We always fix the temperature to 0 when using APIs.

\ifarxiv
\section*{Acknowledgements} 
This paper has received funding from the Swiss
National Science Foundation (SNSF) under the project `How sustainable is sustainable finance? Impact evaluation and automated greenwashing detection' (Grant Agreement No. 100018\_207800).
\else

\fi

\bibliography{custom}

\appendix

\section{Exemplifying Partial Relevance} \label{appendix:partial_relevance}
When labeling whether a document is relevant for a question, there exists a large grey scale of relevance rather than a black-and-white relevant or irrelevant label. Humans can only consistently capture these nuances to a certain extent. The judgment of relevance also depends on the context and the annotator's domain expertise.

Consider for instance the following excerpt of a document:

\begin{lstlisting}[frame=single, basicstyle=\ttfamily\scriptsize, xleftmargin=0pt, numbers=none]
"""
[...] Implement Risk Controls: Integrated Management System (IMS): The K&S Integrated Management System (IMS), which has been implemented at our six major design and manufacturing sites, is certified under the corporate ISO 9001:2015, ISO 14001:2015 and ISO 45001:2018 certifications. Our integrated Quality, Environmental and Occupational Health & Safety (QEHS) Management System enables the achievement of harmonized K&S worldwide objectives.
"""
\end{lstlisting}

Furthermore, conclude as to whether this is relevant to answer the following query and definition:

\begin{lstlisting}[frame=single, basicstyle=\ttfamily\scriptsize, xleftmargin=0pt, numbers=none]
"""
Meaning of the question: The question "What processes does the organization use to identify and assess climate-related risks?" is asking for information about the specific methods, tools, or strategies that a company employs to recognize and evaluate the potential risks to its operations, financial performance, and overall sustainability that are associated with climate change. This includes understanding how the organization anticipates, quantifies, and plans for the impacts of climate-related phenomena such as extreme weather events, long-term shifts in climate patterns, and regulatory changes aimed at mitigating climate change.

Examples of information that the question is looking for:
1. The use of climate risk assessment tools or software that helps in modeling and predicting potential impacts of climate change on the organization's operations.
2. Engagement with external consultants or experts specializing in climate science [...]
"""
\end{lstlisting}

The query clearly looks for processes to identify and assess risks associated with climate change. Example 1. states that "climate risk assessment tools" are relevant. The paragraph states that the Integrated Management System serves to identify risks including environmental risks. In sustainability matters, climate change and environmental topics often fall under the same umbrella. Thus, yes, the paragraph is relevant for the question addressing a certified process to manage climate risks. However, also contrary arguments can be considered. We don't exactly know whether environmental and climate topics are viewed interchangeably. An expert may know clear differentiating factors between environmental and climate matters (e.g., not all environmental problems like water pollution affect the climate). Furthermore, the environmental management system is rather a minor note in this paragraph. Additionally, it seems that, although it is a general risk management system, the "Quality, Environmental and Occupational Health \& Safety (QEHS) Management System" is rather used to achieve worldwide objectives for the company. Would you deem this relevant if it was the only information obtained for a company? And what if there are fifteen more documents that are clearly relevant? How would it be labeled then? It is possible to go to lengths and depending on which expert level or context a labeler holds. In a binary relevant/irrelevant setting, both labels would be partially wrong. The reason lies in the fact that when asking whether this document is relevant to the question, the answer is "partially right".

\section{Creation of the Relevance Definition} \label{appendix:QuestionBackground}
\cref{fig:background_creation_prompt} shows the prompt template for the creation of the query relevance definition. We ask the model to produce a short definition on which the model should rely. Additionally, we ask the model to produce a list of examples. This structure should align with the manner an expert implicity or explicitly approaches the annotation task of labeling relevance. A definition alone would have the shortcoming that it only incorporated generic know-how. Complementing it with examples gives the expert the flexibility to extend the meaning of the terms in exemplified form. For a demonstration of the output, see \cref{tab:generic_specific_example}.

\begin{figure}[ht]
\begin{lstlisting}[frame=single, basicstyle=\ttfamily\scriptsize, xleftmargin=0pt, numbers=none]
"""
An analyst posts a <question> about a climate report. Your task is to explain the <question> in the context of climate reporting. Please first explain the meaning of the <question>, i.e., the meaning of the question itself and the concepts mentioned. And then give a list of examples, showing what information from the climate report the analyst is looking for by posting this <question>.

For <the question's meaning>, please start by repeating the question in the following format:
'''
The question "<question>" is asking for information about [...]
'''

For the <list of example information that the question is looking for>, follow the following example in terms of format:
---
[...]
3. Initiatives aimed at creating new job opportunities in the green economy within the company or in the broader community.
4. Policies or practices in place to ensure that the transition to sustainability is inclusive, considering gender, race, and economic status.
[...]
---

Here is the question:
<question>: ""{question}""

Format your reply in the following template and keep your answer concise:

Meaning of the question: <the question's meaning>
Examples of information that the question is looking for: <list of example information that the question is looking for>"""
\end{lstlisting}
\caption{Prompt for generating a query relevance definition.}
\label{fig:background_creation_prompt}
\end{figure}

\section{Metrics Computation Details} \label{appendix:metrics_details}
In this project, we use Scikit-Learn (version 1.2.2) to compute AUROC, average precision scores, Brier scores, and F1 scores. We employ rank\_eval (version 0.1.3) to compute nDCG and MAP scores, and Scipy.stats to compute Kendall's $\tau$. For nDCG, relevant scores 0.5 and 1 are assigned to partially relevant and relevant documents correspondingly. When averaging Calibration metrics, we average AUROC with $1-$ ECE and $1-$ Brier Score to keep the trend consistent.

\section{ChatReportRetrieve Data Preprocessing}
\label{appendix:document_parsing}

\myparagraph{Data Sampling}: We sample 31 questions and 80 reports from this application. Importantly, ChatReportRetrieve differs from general QA dataset and focuses on integrative queries which usually ask for broad relevant information. Therefore, with controlled annotation budget (i.e., number of (query, document) pairs), we are prone to have fewer (but representative) queries and more documents for each query. The queries are strategically sampled to ensure representativeness and diversity. Specifically, 11 queries are the core queries used in ChatReport, 
which cover essential topics of sustainability disclosure. 20 questions are selected from users' customized questions posed to the ChatReport tool. Climate reports are sampled randomly from openly accessible user submissions\footnote{See \url{https://github.com/EdisonNi-hku/chatreport}.}  
Finally, we prompt GPT-4 to draft relevance definitions for all queries (see \cref{appendix:QuestionBackground}).

\begin{figure}[t]
    \centering
	\includegraphics[width=\columnwidth]{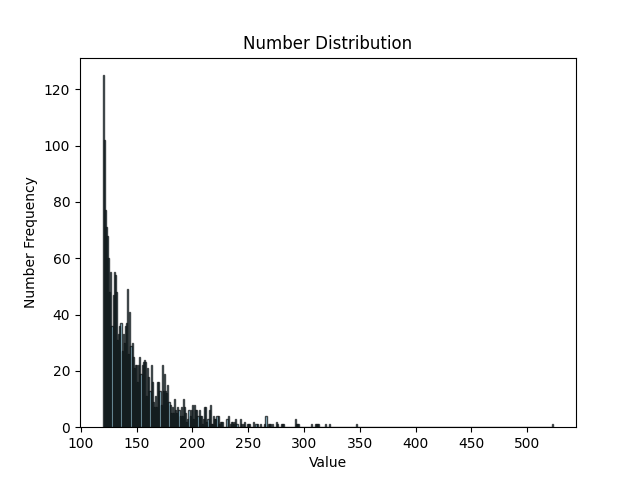}
 \vspace{-2em}
	\caption{Distribution of chunk length after being extracted from climate reports and concatenation.}
	\label{fig: chunks_distribution}
 \vspace{-1em}
\end{figure}

\myparagraph{PDF Parsing}: We use IBM deepsearch parser \citep{deepsearch} to parse corporate reports into chunks. For chunks shorther than 120 tokens, we concatenate them with adjacent chunks to form chunks longer than 120. Figure \cref{fig: chunks_distribution} shows the formatted chunks length distribution. 

\myparagraph{Train-Test Split}: We split the questions into 11 for testing and 20 for training. Similarly, we split the reports into 30 for testing and 50 for training. This ensures the evaluation on unseen queries and reports. For each query, we randomly sample 60 documents -- 30 each from in the top-5 and outside the top-5 (using OpenAI text-embedding-3-small as the dense retriever). 
Ultimately, (query, document) pairs in training split are used to create training data with relevance label and confidence score predictions (details in \cref{sec:prompt_selection}). Data points in the test split are passed to human annotation.
We use 31 climate-related queries for data sampling among 80 climate reports. For the separation of the train dataset and test dataset, these queries are classified into 3 categories: vague, specific, and TCFD. Within each category, queries are split randomly for training and testing, resulting in 20 queries for the train set and 11 queries for the test set. Meanwhile, the 80 climate reports are randomly split into 50 for train and 30 for test.



\section{Expert Annotation Process} \label{appendix:expert_annotation}
As described in \cref{appendix:document_parsing}, the data is obtained from real climate reports and split into chunks of around 150 words with the IBM deepsearch parser \citep{deepsearch}. \cref{tab:stats_test_data} shows an overview of statistical properties of the number of words in test set data.

\begin{table*}[]
\centering
\begin{tabular}{cccccccc}
\hline
\textbf{}      & \multicolumn{7}{c}{\textbf{Number of words per document}}                                                  \\ \hline
\textbf{Dataset Size} & \textbf{Mean} & \textbf{Std} & \textbf{Min} & \textbf{25\%} & \textbf{50\%} & \textbf{75\%} & \textbf{Max} \\ \hline
660            & 150           & 28.5         & 107          & 131           & 143           & 162           & 318          \\ \hline
\end{tabular}
\caption{Statistical properties of the number of words in ChatReportRetrieve test set data.}
\label{tab:stats_test_data}
\end{table*}

Then, we form a group of three expert annotators. The expert annotators comprise one graduate and one PhD student working in NLP for climate change. These two experts label the entire dataset with three labels: the document is relevant, partially relevant, or not relevant for the query including the definition. Following a simple annotation guideline:
\begin{itemize}
    \item Please first carefully read the provided relevance definition to understand what the question is looking for. The definition consists of a question explanation and examples of relevant information.
    \item If a paragraph clearly falls into the definition of relevance, i.e., explicitly mentioned by the question explanation or examples, please annotate relevant.
    \item If the paragraph is not explicitly covered by the definition but you think it somehow helps answering the question. Please annotate partially relevant.
    \item Otherwise please annotate irrelevant.
\end{itemize}

Additionally, one PhD student focusing on climate change and sustainability research serves as a subject-matter meta-annotator to resolve conflicts or investigate cases where both labelers arise at the label partially.

Comparing the two base annotators in the setup, we can calculate inter-annotator agreement. The Cohen's kappa between the two labelers is 0.683 (substantial agreement). We also calculate annotators' agreement on partial relevance. The Cohen's Kappa turns out to be 0.129, suggesting that there are uncertainty and subjectivity associated with partial labels.

Besides the relevance, we also obtain an uncertainty label whenever there is strong disagreement (co-existence of relevance and irrelevance labels) or agreement on partial relevance (two or more annotators agree on partial relevance), the data point is labeled as uncertain. There are 103 (557) uncertain (certain) (query, document) pairs in the dataset.

Finally, the third expert annotator resolves the existing conflicts in the dataset. This results in a label distribution of \cref{tab:lab_dist_test}. It becomes apparent that the majority of documents are not relevant while still a significant number is labeled as partially relevant and relevant.

\begin{table}[]
\centering
\begin{tabular}{cc}
\hline
\textbf{Label} & \textbf{Occurance} \\ \hline
Relevant            & 121                \\
Partially      & 65                 \\
Not Relevant             & 474                \\ \hline
\end{tabular}
\caption{Label distribution in the ChatReportRetrieve testset.}
\label{tab:lab_dist_test}
\end{table}

\section{LLM Fine-Tuning Settings} \label{appendix:ft_detail}
We use the default QLoRA hyperparameter settings \footnote{https://github.com/jondurbin/qlora}, namely, an effective batch size of 32, a lora r of 64, a lora alpha of 16, a warmup ratio of 0.03, a constant learning rate scheduler, a learning rate of 0.0002, an Adam beta2 of 0.999, a max gradient norm of 0.3, a LoRA dropout of 0.1, 0 weight decay, a source max length of 2048, and a target max length of 512. We use LoRA module on all linear layers. All fine-tunings last 2 epochs.

All experiments are conducted on two clusters, one with 4 V100 GPUs and the other with 4 A100 (80G) GPUs. 1 GPU hour is used per fine-tuning.

\section{DIRAS Prompt Template $\mathcal{P}$} \label{appendix:diras_prompt}
\cref{fig:diras_prompt} shows the full prompt DIRAS prompt template for the Chain-of-Thought setup. The non-CoT setup just excludes the ``[Reason]: ..."" part of the prompt. 

\begin{figure}[ht]
\begin{lstlisting}[frame=single, basicstyle=\ttfamily\scriptsize, xleftmargin=0pt, numbers=none]
You are a helpful assistant who assists human analysts in identifying useful information within climate reports for their analysis.

You will be provided with a <question> the analyst seeks to answer, a <paragraph> extracted from a lengthy report, and <background_information> that explains the <question>. <background_information> first explains the <question> and then raises examples to help you to better understand the <question>. Your job is to assess whether the <paragraph> is useful in answering the <question>.

<background_information>: "{background_information}"
<question>: "{question}"
<paragraph>: "{paragraph_chunk}"


Is <paragraph> helpful for answering <question>? Note that the <paragraph> can be helpful even it only addresses part of the <question> without fully answering it. Provide your best guess for this question and your confidence that the guess is correct. Reply in the following format:
[Reason]: <Reason why and how the paragraph is helpful or not helpful for answering the question. Clearly indicate your stance.>
[Guess]: <Your most likely guess, should be one of "Yes" or "No".>
[Confidence]: <Give your honest confidence score between 0.0 and 1.0 about the correctness of your guess. 0 means your previous guess is very likely to be wrong, and 1 means you are very confident about the guess.>
\end{lstlisting}
\caption{Full DIRAS Chain-of-Thought prompt for LLMs predicting relevance labels and calibrating.}
\label{fig:diras_prompt}
\end{figure}

\section{Alternative Prompts} \label{appendix:prompt_selection}
\cref{fig:diras_prompt}, \cref{fig:diras_prompt_prob}, and \cref{fig:diras_prompt_no_e} show the alternative prompts with which we experimented.

\begin{figure}[ht]
\begin{lstlisting}[frame=single, basicstyle=\ttfamily\scriptsize, xleftmargin=0pt, numbers=none]
{Same task description and inputs}

Is <paragraph> helpful for answering <question>? Note that the <paragraph> can be helpful even it only addresses part of the <question> without fully answering it. Provide your best guess for this question and the probability that the <paragraph> is helpful. Reply in the following format:
[Reason]: <Reason why and how the paragraph is helpful or not helpful for answering the question. Clearly indicate your stance.>
[Guess]: <Your most likely guess, should be one of "Yes" or "No".>
[Probability Helpful]: <The probability between 0.0 and 1.0 that the <paragraph> is helpful to the <question>. 0.0 is completely unhelpful and 1.0 is completely helpful.>
\end{lstlisting}
\caption{Output requirements for the alternative prompt setting $\mathcal{P}_{prob}$. Task description and input are the same as \cref{fig:diras_prompt}.}
\label{fig:diras_prompt_prob}
\end{figure}

\begin{figure}[ht]
\begin{lstlisting}[frame=single, basicstyle=\ttfamily\scriptsize, xleftmargin=0pt, numbers=none]
You will be provided with a <question> the analyst seeks to answer, and a <paragraph> extracted from a lengthy report. Your job is to assess whether the <paragraph> is useful in answering the <question>.

<question>: "{question}"
<paragraph>: "{paragraph_chunk}"

{Same output requirements}
\end{lstlisting}
\caption{Task description and input part for the alternative prompt setting $\mathcal{P}_{w\text{/}o\_e}$. Output requirements are the same as \cref{fig:diras_prompt}.}
\label{fig:diras_prompt_no_e}
\end{figure}

\begin{figure}[ht]
\begin{lstlisting}[frame=single, basicstyle=\ttfamily\scriptsize, xleftmargin=0pt, numbers=none]
<|system|>
You are RankLLM, an intelligent assistant that can rank passages based on their relevancy to the query.
<|user|>
I will provide you with {num} passages, each indicated by a numerical identifier [].
Rank the passages based on their relevance to the search query: {query}.
{passages}
Search Query: {query}.
Rank the {num} passages above based on their relevance to the search query. All the passages should be included and listed using identifiers, in descending order of relevance. The output format should be [] > [], e.g., [4] > [2]. Only respond with the ranking results, do not say any word or explain.
\end{lstlisting}
\caption{We use exactly the same listwise ranking prompt as \citet{sun-etal-2023-chatgpt} and \citet{pradeep2023rankzephyr}. Both system and user prompts are presented in this figure.}
\label{fig:list_original}
\end{figure}

\begin{figure}[ht]
\begin{lstlisting}[frame=single, basicstyle=\ttfamily\scriptsize, xleftmargin=0pt, numbers=none]
<|system|>
You are RankLLM, an intelligent assistant that can rank passages based on their relevancy to the query.
<|user|>
I will provide you with {num} passages, each indicated by a numerical identifier [].
Rank the passages based on their relevance to the search query: {query}.
{passages}
Search Query: {query}.

Here are some background information that explains the query: {relevance_definition}

Rank the {num} passages above based on their relevance to the search query. All the passages should be included and listed using identifiers, in descending order of relevance. The output format should be [] > [], e.g., [4] > [2]. Only respond with the ranking results, do not say any word or explain.
\end{lstlisting}
\caption{Listwise prompt with an extra input of explicit definition.}
\label{fig:list_definition}
\end{figure}

\section{Creation of the Improved Relevance Definitions} \label{appendix:specific_query_definition}
\cref{fig:specific_background_creation_prompt} shows the prompt for the creation process of the improved relevance definitions. Following the procedure in \citet{schimanski2024climretrieve}, we make use of the text parts labeled as relevant. There exists a relevance score from 1-3 where 1 signals the least and 3 is most relevant. Similar to the base setup for the experiments in \citet{schimanski2024climretrieve}, we use the text samples with a score of 2 or higher to create the improved relevance definition. We include relevant text samples in the prompt for creating the relevance definitions to obtain improved definitions.
The logic behind this definition creation is that we assume we know the mental model of ClimRetrieve human analysts, and thus know what information is relevant before annotation. This is common in corporate report analysis where experts will have fixed concepts in their heads, maybe even inspired by prior search processes. 

Plugging the examples into the prompt results in a set of improved relevance definitions. When comparing these relevance definitions to the generic ones, it becomes apparent that GPT-4 already incorporated the majority of the concepts that the experts were looking for. Therefore, the adjustment of the relevance definition is visible but rather subtle. One example is displayed in \cref{tab:generic_specific_example}. While the meaning of the question remains rather static, there are nuanced differences in the examples that guide the relevance labeling.

\begin{figure}[ht]
\begin{lstlisting}[frame=single, basicstyle=\ttfamily\scriptsize, xleftmargin=0pt, numbers=none]
"""
An analyst posts a <question> about a climate report. Your task is to explain the <question> in the context of climate reporting. Please first explain the meaning of the <question>, i.e., meaning of the question itself and the concepts mentioned. And then give a list of examples, showing what information from the climate report the analyst is looking for by posting this <question>.

For <the question's meaning>, please start by repeating the question in the following format:
'''
The question "<question>" is asking for information about [...]
'''

For the <list of example information that the question is looking for>, following the following example in terms of format:
---
[...]
3. Initiatives aimed at creating new job opportunities in the green economy within the company or in the broader community.
4. Policies or practices in place to ensure that the transition to sustainability is inclusive, considering gender, race, and economic status.
[...]
---

Here is the question:
<question>: ""{question}""

Additionally, here is a <list of question-relevant example information> that an expert human labler annotated. Please keep these examples in mind when answering:
--- [BEGIN <list of question-relevant example information>]
{examples}
--- [END <list of question-relevant example information>]

Format your reply in the following template and keep your answer concise:

Meaning of the question: <the question's meaning>
Examples of information that the question is looking for: <list of example information that the question is looking for>"""
\end{lstlisting}
\caption{Prompt Template enforcing structured output with the inclusion of examples.}
\label{fig:specific_background_creation_prompt}
\end{figure}

\begin{table*}[ht]
\small
\centering
\begin{tabular}{>{\raggedright\arraybackslash}m{0.1\textwidth}m{0.41\textwidth}m{0.39\textwidth}}
\hline
Question                                                                                                   & Generic Definition                                                                                                                                                                                                                                                                                                                                                                                                                                                                                                                                                                                                                                                                                                                                                                                                                                                                                                                                                                                                                                                                                                                                                                                                                                                                                                                                                                                                                                                                                                                                                                                                                                                                                                                                                                                                                                                                                                                     & Expert-informed Definition                                                                                                                                                                                                                                                                                                                                                                                                                                                                                                                                                                                                                                                                                                                                                                                                                                                                                                                                                                                                                                                                                                                                                                                                                                                                                                                                                                                                                                                                                                                                                                                                                                                                                                                                                                                                                                                                                                                                                                                                                                                                                                                                                                                                                                                                                                                                                                                                                                                                                                                                                                                                                                                                              \\ \hline
Does the company provide definitions for climate change adaptation?
 & \begin{lstlisting}[frame=single, basicstyle=\ttfamily\scriptsize, xleftmargin=0pt, numbers=none]
"Meaning of the question: The question ""Does the company provide definitions for climate change adaptation?"" is asking for information about whether the company has clearly outlined what it understands by the term ""climate change adaptation"" within its climate reporting. This involves understanding how the company perceives the process of adjusting to actual or expected climate change and its effects, as well as how it plans to manage or mitigate harm or exploit beneficial opportunities.

Examples of information that the question is looking for:
1. A clear definition of climate change adaptation as understood and applied by the company in its operations and planning.
2. Descriptions of the specific impacts of climate change that the company is adapting to, such as increased frequency of extreme weather events, rising sea levels, or changing agricultural conditions.
3. Examples of measures the company is taking to adapt to these impacts, such as investing in flood defenses, altering supply chain logistics, or developing drought-resistant crops.
4. Information on how the company's definition of climate change adaptation aligns with or differs from standard definitions provided by authoritative bodies like the Intergovernmental Panel on Climate Change (IPCC) or the United Nations Framework Convention on Climate Change (UNFCCC).
5. Details on any collaborative efforts with governments, NGOs, or other corporations to address climate change adaptation collectively.
6. Insights into how the company measures or evaluates the effectiveness of its adaptation strategies."
\end{lstlisting} & \begin{lstlisting}[frame=single, basicstyle=\ttfamily\scriptsize, xleftmargin=0pt, numbers=none]
"Meaning of the question: The question ""Does the company provide definitions for climate change adaptation?"" is asking for information about whether the company has clearly defined what it means by climate change adaptation within its climate reporting. This involves understanding how the company interprets climate change adaptation, including any specific strategies, initiatives, or policies it has in place to adjust to current or expected future climate change impacts.

Examples of information that the question is looking for:
1. Descriptions of how the company defines climate change adaptation in the context of its operations and strategic planning.
2. Examples of specific adaptation measures the company has implemented or plans to implement, such as enhancing infrastructure resilience, diversifying water sources, or adjusting agricultural practices.
3. Information on how the company's definition of climate change adaptation aligns with or diverges from standard definitions provided by environmental organizations or regulatory bodies.
4. Details on how the company assesses and integrates climate change risks and opportunities into its investment decision-making processes, focusing on adaptation.
5. Statements on the company's involvement in partnerships or alliances aimed at promoting climate change adaptation and resilience, indicating a collaborative approach to defining and addressing adaptation needs."
\end{lstlisting} \\ \hline
\end{tabular}

\caption{Example of a generic and expert-informed relevance definition for a question.}
\label{tab:generic_specific_example}
\end{table*}

\section{MAP and nDCG Scores for Different Relevance Definitions} \label{appendix:map_nDCG_qeustion_definition}
In the improved definition experiment, we compare two settings. First, we compare the predictions on the 595 relevant-only (query, document) pairs. This is a replication of the setting in \cref{sec:climre_veri}. Since we do not have non-relevant samples, we can only compare the nDCG. \cref{table:comp_gen_spec_relonly} shows the results. It becomes apparent that only for the general nDCG score, the improved query performs better. For the nDCG@5, and nDCG@10, the best-performing model remains with the generic prompt. The picture turns again when widening to nDCG@15. This could be a result of the definition creation. We use examples of relevance labels 2 and 3 to create the improved definition with GPT-4. Thus, we implicitly equalize relevance 2 and 3 in importance. This means we are likely less effective in differentiating between 2 and 3. This could explain the lower results at lower k's where differentiating between 2 and 3 is important v.s. the overall nDCG where differentiating between 1 and 2/3 plays a more important role.

This intuition is reinforced by the second setting, comparing the predictions on all 43K (query, document) pairs. In this setting, we also calculate the relevance for a large amount of non-relevant pairs. As \cref{table:comp_gen_spec_all} shows, the expert-informed definition now seems effective, especially when comparing MAP. MAP is agnostic to the actual degree of relevance and rather just differentiates between relevant and not relevant. Thus, the clearly higher MAP scores show that the expert-informed definition helps in differentiating between the non-relevant pairs where the definition is not meant for vs. those the definition was created with and for. This indicates that our approach is indeed sensitive to adjusting the relevance definitions.

\begin{table*}[]
\centering
\small
\begin{tabular}{lcccc}
\hline
\textbf{Setting}     & \textbf{nDCG}  & \textbf{nDCG@5} & \textbf{nDCG@10} & \textbf{nDCG@15} \\ \hline
Llama3-Ask$_{generic}$  & \underline{77.23}    & \textbf{67.60}  & \textbf{66.18}   & \textbf{67.57}   \\
Llama3-Tok$_{generic}$  & 76.55          & \underline{67.20}     & \underline{66.23}      & 65.83            \\
Llama3-Ask$_{informed}$ & 76.52          & 63.24           & 65.69            & 66.39            \\
Llama3-Tok$_{informed}$ & \textbf{77.41} & 65.95           & 65.06            & \underline{66.91}      \\ \hline
\end{tabular}
\caption{Comparison of using the generic and the expert-informed relevance definitions for ranking \textbf{relevant only} ClimRetrieve (query, document) pairs.}
\label{table:comp_gen_spec_relonly}
\end{table*}

\begin{table*}[h!]
\centering
\small
\begin{tabular}{lcccccccc}
\hline
\textbf{Setting} & \textbf{nDCG} & \textbf{nDCG@5} & \textbf{nDCG@10} & \textbf{nDCG@15} & \textbf{MAP} & \textbf{MAP@5} & \textbf{MAP@10} & \textbf{MAP@15} \\
\hline
Llama3-Ask$_{generic}$ & 29.95 & 18.67 & 21.71 & 23.38 & 26.51 & 17.86 & 21.21 & 22.75 \\
Llama3-Tok$_{generic}$ & \underline{31.17} & \underline{20.35} & \underline{23.21} & \underline{25.17} & 28.73 & 19.58 & 23.15 & 25.05 \\
Llama3-Ask$_{informed}$ & 30.89 & 19.01 & 22.82 & 24.89 & \underline{29.31} & \underline{20.02} & \underline{23.60} & \underline{25.56} \\
Llama3-Tok$_{informed}$ & \textbf{32.53} & \textbf{21.47} & \textbf{24.99} & \textbf{26.92} & \textbf{32.65} & \textbf{22.97} & \textbf{27.20} & \textbf{28.77} \\
\hline
\end{tabular}
\caption{Comparison of using the generic and the expert-informed relevance definitions for ranking \textbf{all} ClimRetrieve (query, document) pairs.}
\label{table:comp_gen_spec_all}
\end{table*}

\section{Embedding Fine-Tuning} \label{appendix:embedding_ft}
We follow the official fine-tuning example\footnote{\url{https://github.com/FlagOpen/FlagEmbedding/tree/master/examples/finetune}} of \citep{chen2024bge} to fine-tune the embedding models. The models are fine-tuned on all annotated (query, document) pairs in ChatReportRetrieve test set for 10 epochs, with a batch size of 4. Other hyperparameters are the same as the official example.

\section{Hand-Checking DIRAS Model Disagreements with ClimRetrieve} \label{appendix:error_analysis}

We want to check samples where our student model Llama3-Tok’s prediction differs from the relevance label in ClimRetrieve. We sample equally from those samples where Llama3-Tok indicated relevance and ClimRetrieve does not and vice versa. While our focus lies on those documents that were not annotated in ClimRetrieve (i.e. irrelevant ones), the dataset also allows us to investigate how our model performs in the edge case of when a ClimRetrieve annotator deems relevance and Llama3-Tok does not.

On the samples that were labeled as irrelevant in ClimRetrieve, we find that our student model Llama3-Tok model is effective in mitigating annotation selection bias and therefore incorporates the perspective of IR recall (see \cref{tab:climeretrieve_disagreement}).

However, it is interesting that our student model Llama3-Tok successfully overrules human decisions. We attribute this to the fact that our created relevance definitions might differ from the mental model of the human annotator in ClimeRetrieve. Thus, humans in ClimRetrieve might have been consistent with their own mental model. For us, however, it is more important and reaffirming to observe that Llama3-Tok is consistent with its own, explicit relevance definitions. 

We can view the (query, definition, document) pair in \cref{fig:DIRASvsClimRetrieve} as an example. When analyzing the query "Do the environmental/sustainability targets set by the company reference external climate change adaptation goals/targets?", the ClimRetrieve labeler interpreted the question broader, i.e., deeming this as relevant: "As a global technology leader, we are also committed to helping build the enabling societal conditions that will support a net zero economy.". However, for our student model, it is in line with the definition to assign a "not relevant" label. There is no explicit standard mentioned in the document.

\begin{figure}
\begin{lstlisting}[frame=single, basicstyle=\ttfamily\scriptsize, xleftmargin=0pt, numbers=none]
QUERY: "Do the environmental/sustainability targets set by the company reference external climate change adaptation goals/targets?"

QUERY DEFINITION: Meaning of the question: The question "Do the environmental/sustainability targets set by the company reference external climate change adaptation goals/targets?" is asking for information about whether the company's stated goals or objectives for environmental sustainability or climate change mitigation are aligned with, or make reference to, established external goals or targets. These external references could include international agreements, national policies, or standards set by recognized organizations focused on climate change and sustainability.

Examples of information that the question is looking for:
1. In line with our commitment to the Net-Zero Banking Alliance (NZBA) [...]

DOCUMENT: Enabling a more sustainable world Microsoft's actions alone will not solve the climate crisis. As a global technology leader, we are also committed to helping build the enabling societal conditions that will support a net zero economy. We're focused on accelerating the availability of new climate technologies, strengthening our climate policy agenda, helping to develop a more reliable and interoperable carbon accounting system, advocating for skilling programs to expand the green workforce, and working to enable a just energy transition.
\end{lstlisting}
\caption{Example for which  DIRAS Llama3-Tok assigns "not relevant" and ClimRetrieve assigned "relevant". This example shows that while DIRAS Llama3-Tok might differ with the opinion of ClimRetrieve's annotator, it is consistent with its own definition.}
\label{fig:DIRASvsClimRetrieve}
\end{figure}

\section{Implementation Details of Experiments on QA Datasets} \label{appendix:details_qa}
\myparagraph{ALCE Data}: We obtain ELI5, ASQA, and QAMPARI from ALCE \citep{gao2023enablinglargelanguagemodels}, where they parse the original open-domain QAs into RAG forms\footnote{Data files in \url{https://github.com/princeton-nlp/ALCE}}. For each question, ALCE annotates 5 documents as oracle based on retrieval recall and reference answers, which are used as context relevance labels in our experiment. For each dataset, we randomly sample 100 queries to construct DIRAS training data, following the process in \cref{fig:overview}. Top-5 is selected for balanced sampling, thus resulting in 1000 (query, document) pairs for training. We sample 50 questions for test data, and include all (query, document) pairs for them, resulting in 5K (query, document) pairs for each dataset.

\myparagraph{RAG-Bench Data}: RAG-Bench classifies RAG sources into four types: (A) relevant and with answers, (B) relevant topic but without answers, (C) irrelevant topic, and (D) with counterfactual answers. We find (B) addresses partial relevance that DIRAS cares about. Therefore, we leverage its dev set for training and test set for testing, where (A) and (D) become relevant documents, and (B) and (C) are used as irrelevant ones.

\myparagraph{Disagreement Sampling for \cref{tab:qa_disagreement}}: We sample 200 disagreed annotations four each dataset, 50 samples from each confidence range: Conf$<90$, $90<$Conf$<95$, $95<$Conf$<98$, and $98<$Conf$<100$ to balancedly cover different confidence scores (these bins are of similar size). 

\begin{figure}[ht]
\begin{lstlisting}[frame=single, basicstyle=\ttfamily\scriptsize, xleftmargin=0pt, numbers=none]
You will be provided with a <question> and a <paragraph>.  Your job is to assess whether the <paragraph> is useful in answering the <question>, given the <background_information> defining what is useful.

<background_information>: "The <paragraph> is useful only if some of its content directly answer the <question> or at least a part of the <question>. Content with relevant topic but without direct answers are not useful."
<question>: "{question}"
<paragraph>: "{paragraph_chunk}"


Is <paragraph> useful for answering <question>? Provide your best guess and your confidence that the guess is correct. Reply in the following format:
[Reason]: <Reason why and how the paragraph is helpful or not helpful for answering the question. Clearly indicate your stance.>
[Guess]: <Your most likely guess, should be one of "Yes" or "No".>
[Confidence]: <Give your honest confidence score between 0.0 and 1.0 about the correctness of your guess. 0 means your previous guess is very likely to be wrong, and 1 means you are very confident about the guess.>
\end{lstlisting}
\caption{The DIRAS prompt for QA experiments.}
\label{fig:qa_prompt}
\end{figure}
\end{document}